\documentclass[aps,prl,superscriptaddress,citeautoscript,twocolumn]{revtex4-1}

\usepackage{graphicx}
\usepackage{dcolumn}
\usepackage{amsmath,mathrsfs,amssymb,amsthm}
\usepackage{hhline}
\usepackage{wasysym,enumerate,color}
\usepackage{epstopdf}
\usepackage{verbatim}
\usepackage{hyperref}
\usepackage{color}
\usepackage{ulem} % needed for \scrap-command

\newcommand{\bea}{\begin{eqnarray*}}
\newcommand{\eea}{\end{eqnarray*}}
\newcommand{\bne}{\begin{equation*}}
\newcommand{\ede}{\end{equation*}}

\newcommand{\bnen}{\begin{equation}}
\newcommand{\eden}{\end{equation}}
\newcommand{\bean}{\begin{eqnarray}}
\newcommand{\eean}{\end{eqnarray}}
\newcommand{\bnsn}{\begin{subequations}}
\newcommand{\edsn}{\end{subequations}}

\newcommand{\bna}{\begin{array}}
\newcommand{\eda}{\end{array}}
\newcommand{\bnm}{\begin{enumerate}}
\newcommand{\edm}{\end{enumerate}}

\newcommand{\bk}{\mathbf k}
\newcommand{\bS}{\mathbf S}
\newcommand{\bs}{\boldsymbol \sigma}
\newcommand{\bx}{\mathbf x}
\newcommand{\s}{\sigma}

\definecolor{darkgreen}{rgb}{0,0.5,0}
\definecolor{purple}{rgb}{0.35,0,0.35}
\definecolor{orange}{rgb}{1,0.5,0}
\definecolor{darkred}{rgb}{.7,0,0}
\definecolor{darkblue}{rgb}{0,0,.3}
\definecolor{grey}{rgb}{.6,.6,.6}
\definecolor{dimgreen}{rgb}{0.2,0.6,0.1}
\hypersetup{colorlinks,linkcolor=blue,urlcolor=blue,citecolor=blue}

\newcommand{\GT}{$G_{\rm T}$ }

\newcommand{\gB}{$g_{\rm B}$ }
\newcommand{\IT}{$I_{\rm T}$ }
\newcommand{\VP}{$V_{\rm P}$ }
\newcommand{\VB}{$V_{\rm B}$ }
\newcommand{\GS}{$\Gamma_{\rm S}$ }

\newcommand{\db}{$|D\rangle$}
\newcommand{\sg}{$|S\rangle$}

\begin{document}

\title{Large spatial extension of the zero-energy Yu-Shiba-Rusinov state in magnetic field}

\author{Zolt\'an~Scher\"ubl}
\affiliation{Department of Physics, Budapest University of Technology and Economics and MTA-BME "Momentum" Nanoelectronics Research Group, H-1111 Budapest, Budafoki \'ut 8., Hungary}
\author{Gerg\H{o}~F\"ul\"op}
\affiliation{Department of Physics, Budapest University of Technology and Economics and MTA-BME "Momentum" Nanoelectronics Research Group, H-1111 Budapest, Budafoki \'ut 8., Hungary}
\affiliation{Department of Physics, University of Basel, Klingelbergstrasse 82, CH-4056 Basel, Switzerland}
\author{C\u at\u alin~Pa\c scu~Moca}
\affiliation{Department of Theoretical Physics and BME-MTA Exotic Quantum Phases Research Group, Budapest University of Technology and Economics, 1521 Budapest, Hungary}
\affiliation{Department of Physics, University of Oradea, 410087, Oradea, Romania}
\author{J\"org Gramich}
\affiliation{Department of Physics, University of Basel, Klingelbergstrasse 82, CH-4056 Basel, Switzerland}
\author{Andreas~Baumgartner}
\affiliation{Department of Physics, University of Basel, Klingelbergstrasse 82, CH-4056 Basel, Switzerland}
\affiliation{Swiss Nanoscience Institute, Klingelbergstrasse 82, Ch-4056 Basel, Switzerland}
\author{P\'eter~Makk}
\affiliation{Department of Physics, Budapest University of Technology and Economics and MTA-BME "Momentum" Nanoelectronics Research Group, H-1111 Budapest, Budafoki \'ut 8., Hungary}
\affiliation{Department of Physics, University of Basel, Klingelbergstrasse 82, CH-4056 Basel, Switzerland}
\author{Tosson~Elalaily}
\affiliation{Department of Physics, Budapest University of Technology and Economics and MTA-BME "Momentum" Nanoelectronics Research Group, H-1111 Budapest, Budafoki \'ut 8., Hungary}
\affiliation{Department of Physics, Faculty of Science, Tanta University, Al-Geish St., 31111 Tanta, Gharbia, Egypt.}
\author{Christian~Sch\"onenberger}
\affiliation{Department of Physics, University of Basel, Klingelbergstrasse 82, CH-4056 Basel, Switzerland}
\affiliation{Swiss Nanoscience Institute, Klingelbergstrasse 82, CH-4056 Basel, Switzerland}
\author{Jesper~Nyg\r{a}rd}
\affiliation{Center for Quantum Devices and Nano-Science Center, Niels Bohr Institute, University of Copenhagen, Universitetsparken 5, DK-2100 Copenhagen, Denmark}
\author{Gergely~Zar\'and}
\affiliation{Exotic Quantum Phases "Momentum" Research Group, Budapest University of Technology and Economics, 
H-1111 Budapest, Hungary}
\author{Szabolcs~Csonka}
\email{csonka@mono.eik.bme.hu}
\affiliation{Department of Physics, Budapest University of Technology and Economics and MTA-BME "Momentum" Nanoelectronics Research Group, H-1111 Budapest, Budafoki \'ut 8., Hungary}

\begin{abstract}
Various promising qubit concepts have been put forward recently based on engineered superconductor (SC) subgap states like Andreev bound states, Majorana zero modes or the Yu-Shiba-Rusinov (Shiba) states. The coupling of these subgap states via a SC strongly depends on their spatial extension and is an essential next step for future quantum technologies. Here we investigate the spatial extension of a Shiba state in a semiconductor quantum dot coupled to a SC for the first time. With detailed transport measurements and numerical renormalization group calculations we find a remarkable more than 50~nm extension of the zero energy Shiba state, much larger than the one observed in very recent scanning tunneling microscopy (STM) measurements. Moreover, we demonstrate that its spatial extension increases substantially in magnetic field.
\end{abstract}

\date{\today}
\maketitle

Superconductor nanostructures are the most advanced platforms for quantum computational architectures thanks to the macroscopic coherent wavefunction and the robust protection by the superconducting gap. Recently, various novel qubit concepts like the Andreev (spin) qubits \cite{PadurariuEPL2012,JanvierScience2015,ParkPRB2017,HaysPRL2018,TosiPRX2019}, Majorana box qubits \cite{LeijnseSemiSciTech2012,AasenPRX2016,Aguado2017}, braiding with Majorana zero modes in a Majorana or a Shiba-chain \cite{SauNatComm2012,ChoyPRB2011,NadjPergePRB2013,BrauneckerPRL2013,PientkaPRB2013,KlinovajaPRL2013,NakosaiPRB2013,VazifehPRL2013,KimPRB2014,NadjPergeScience2014} have been put forward or even implemented. All these qubits are based on their associated sub-gap states such as Andreev bound states \cite{BeenakkerPRL1991}, Majorana zero modes \cite{SauPRL2010,AliceaPRB2010,LutchynPRL2010,OregPRL2010,LeijnsePRB2012,MourikScience2012,NadjPergeScience2014,DengScience2016} or Shiba states \cite{YuActaPhysSin1965,ShibaProgTheorPhys1968,RusinovJETP1969,BalatskyRMP2006}. The Shiba state is formed when a magnetic adatom or its artificial version (quantum dot) is coupled to a superconductor and the localized magnetic moment creates a subgap state by binding an anti-aligned quasiparticle from the superconductor. Depending on the coupling strength between the superconductor and the magnetic moment, the ground state can be either the screened local moment with singlet character, \sg or the unscreened doublet states, \db.

The coupling of these sub-gap states via a superconductor is an essential next step towards 2-qubit operations or state engineering, e.g. an Andreev molecule \cite{PilletArxiv2018,ScherublBJNano2019,KornichArXiv2019} or a Majorana-chain, which consists of series of adatoms or quantum dots interlinked by the superconductor \cite{SauNatComm2012,ChoyPRB2011,NadjPergePRB2013,BrauneckerPRL2013,PientkaPRB2013,KlinovajaPRL2013,NakosaiPRB2013,VazifehPRL2013,KimPRB2014,NadjPergeScience2014,KimSciAdv2018,SteinbrecherNatComm2018,KamlapureNatComm2018}. Obviously, the coupling between such sub-gap states strongly depends on their spatial extension into the superconductor, so it is required for these localized states to extend as much as possible.

So far, the spatial extent and structure of the Shiba states was investigated by STM measurements on magnetic adatoms deposited on the surface of a superconductor \cite{JiPRL2008,ChoiNatComm2017,MenardNatPhys2015,RubyPRL2016} and, interestingly, it revealed that the dimensionality plays a crucial role \cite{MenardNatPhys2015}. In a three dimensional isotropic s-wave superconductor, it was found that the Shiba states decay over a very short distance of the order of $\sim 1$~nm \cite{JiPRL2008,ChoiNatComm2017}, but extends one order of magnitude further, as far as $\sim 10$~nm, if the impurity is placed on the surface of a two-dimensional superconductor \cite{MenardNatPhys2015,MenardNatComm2017}.

In this work, we investigate the spatial extension of the Shiba state formed when an artificial atom is strongly coupled to a superconductor. The Shiba state is observed at a remarkably large distance of more than 50~nm. Furthermore, for the first time we explore the effect of an external magnetic field on the extension of the Shiba state, as it is relevant to access topological superconducting states. Remarkably, with increasing magnetic field, the spatial extension increases significantly further. 

Shiba states were widely studied in two different types of systems: a) in STM measurements, when magnetic particles are deposited on the surface of a superconductor \cite{YazdaniScience1997,JiPRL2008,HatterNatComm2015,MenardNatPhys2015,RubyPRL2016,ChoiNatComm2017,KezilebiekeArXiv2017,CornilsPRL2017,MenardNatComm2017,FarinacciPRL2018,HeinrichProgSurfSci2018,SchneiderArXiv2019}, and b) in nanocircuits, when a quantum dot is attached to the superconductor \cite{BuitelaarPRL2002,EichlerPRL2007,JespersenPRL2007,DeaconPRL2010,DeaconPRB2010,DirksNatPhys2011,KimPRL2013,PilletNatPhys2010,ChangPRL2013,PilletPRB2013,KumarPRB2014,SchindelePRB2014,LeeNatNano2014,JellinggaardPRB2016,LeePRB2017,GramichPRB2017,LiPRB2017,BretheauNatPhys2017,SuNatComm2017,SuPRL2018,SuArXiv2019}.
The STM geometry allows for the spatial mapping of the Shiba state \cite{JiPRL2008,ChoiNatComm2017,MenardNatPhys2015,RubyPRL2016}, but the strength of the coupling between the magnetic adatom and the substrate is mostly determined by the microscopic details and its tuning remains quite challenging \cite{HatterNatComm2015,FarinacciPRL2018}. In contrast, the quantum dot realization enables the tuning the energy of the Shiba state via the level position or the tunnel couplings by using external gate voltages \cite{LeeNatNano2014,JellinggaardPRB2016}. Another advantage of the latter setup is the potential to apply an external magnetic field without stability issues. 
	
\vspace{1cm}
\textbf{Results}
	
\begin{figure}[tb]
	\begin{center}
	\includegraphics[width=\columnwidth]{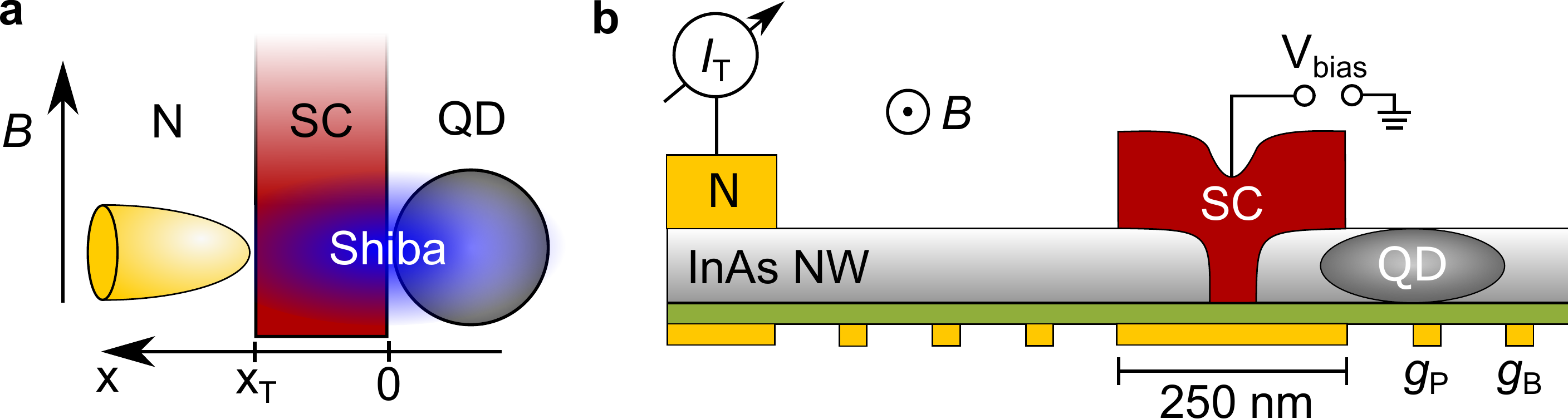}
	\caption{Schematics of the setup. \textbf{a} The normal metal-superconductor-quantum dot (N--SC--QD) setup used in our measurements. The QD is strongly coupled to the s-wave superconductor, giving rise to the Shiba state. The normal lead N is coupled to the superconductor at a finite distance $x_{\text{T}}$, from the dot and acts as a tunnel probe and measures the current as the energy of the Shiba state is tuned. The external magnetic field B is applied in-plane of the wafer. 
	\textbf{b} Cross section of the device. The QD is formed in an InAs nanowire (gray) by applying voltages on the bottom gate electrodes (yellow). The plunger gate $g_{\text{P}}$ controls the level position of the dot, the barrier gate \gB isolates the QD. Tunneling to N is controlled by a series of gates. The width of the superconductor is 250~nm and below the superconductor a segment with length of $50-60$~nm is cut from the nanowire.}
	\label{fig1}
	\end{center}
	\end{figure}

\textbf{Implementation of the Shiba device.}   
In this paper, we implement a combined approach of systems a) and b), where a tunnel probe is attached to a superconductor--quantum dot hybrid. The schematics of the used device are shown on Fig.~\ref{fig1}a. A quantum dot (QD in gray) is strongly coupled to a superconductor (SC in red), leading to the formation of a Shiba state. An additional tunnel electrode (N in yellow) is coupled weakly to the superconductor at a fixed distance from the dot. Applying a small bias between SC and N, the tunnel current, \IT and the corresponding differential conductance, \GT are measured, while the energy of the Shiba state is tuned by the plunger gate $g_{\text{P}}$.

The device is implemented in an InAs semiconducting nanowire (gray), contacted by a 250~nm wide Pb superconducting electrode in the middle (SC in red), and one normal contact (N in yellow) on the left side (see Fig.~\ref{fig1}b). The electron density in the nanowire is tuned by an array of gates fabricated below the nanowire. The nanowire was cut by focused ion beam (FIB) prior to the deposition of the superconducting contact to suppress the direct tunnel coupling between the two arms \cite{FulopPRB2014,FulopPRL2015}. The width of the FIB cut is about $\sim 50-60$~nm. The quantum dot is formed in the right arm of the wire and its level position is tuned by the voltage $V_{\text{P}}$ on the plunger gate $g_{\text{P}}$. The tunnel coupling to the right side can be turned on and off by the barrier gate $g_{\rm B}$. Although not displayed in Fig.~1, there is a normal electrode to the right of $g_{\text{B}}$, which allows us to measure direct transport through the quantum dot (see Methods). The bulk coherence length for Pb is about $\xi_0\simeq 80$~nm, however, in e-beam evaporated layers the elastic mean free path is considerably reduced~\cite{TinkhamPRL1961}, limiting the coherence length to $\xi_0 =20-40$~nm. An in-plane magnetic field $B$ is applied perpendicular to the nanowire axis. Further details on the fabrication and the experimental techniques are presented in the Methods.

\begin{figure*}[tb]
	\begin{center}
	\includegraphics[width=0.8\textwidth]{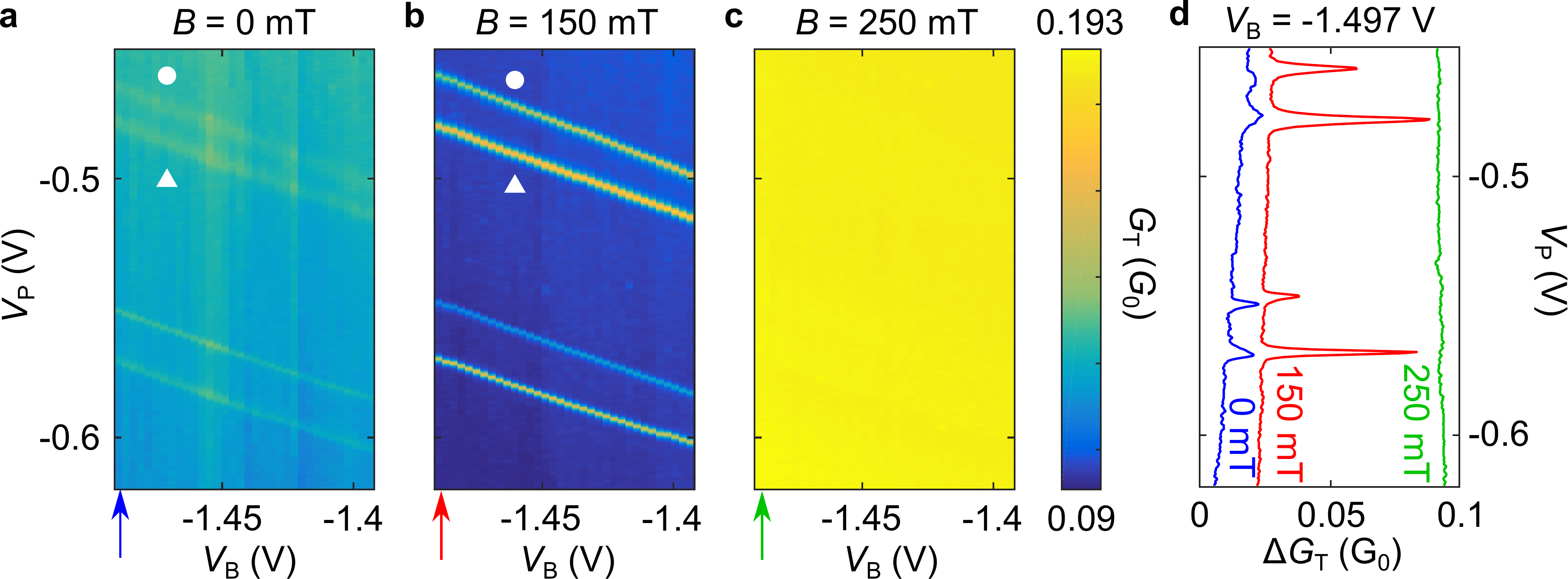}
	\caption{\textbf{Observation of the Shiba state in the tunnel current.} \textbf{a-c} Differential conductance $G_{\rm T}$ for the tunnel probe as the function of the barrier gate voltage $V_{\text{B}}$ and plunger gate voltage $V_{\text{P}}$ for various magnetic fields, as indicated in the panels. The two pair of parallel resonances in the differential conductance map are signatures of the even-odd and odd-even transition of the Shiba state (the upper ones are marked by white circle and triangle)
	{\bf{d}} line-cuts at $V_{\rm B}= -1.497$~V for each value of the magnetic field is represented (indicated by the arrows below panels \textbf{a-c}). The blue line corresponds to $B=0$, the red line to $B=150$~mT and the green line to $B=250$~mT. Since the critical magnetic field is $B_c= 230$~mT, the differential conductance enhancement vanishes for $B>B_c$.}
	\label{fig2}
	\end{center}
	\end{figure*}

\vspace{1cm}
\textbf{Observation of the Shiba state in the tunnel current.}
First we present the results of transport measurements for the strongly coupled superconductor -- quantum dot setup isolated from the rest of the device on the right by $g_{\text{B}}$. The differential conductance of the tunnel probe, \GT is shown in Fig.~\ref{fig2} as function of $V_{\text{B}}$ and \VP for different values of the magnetic field. In the absence of magnetic field (see panel a), two pairs of resonances with enhanced conductance are present on top of a smooth conductance background of about $0.1~G_0$ (with $G_0 = 2e^2/h$ the conductance quantum). The conductance increment along the lines is about $0.005~G_0$. We use the plunger gate voltage \VP to tune the level position and subsequently the charge on the quantum dot, while the barrier gate voltage \VB is used to isolate the dot from the rest of the nanowire on its right side. The barrier gate $g_{\text{B}}$ has a cross capacitance to the dot, resulting a tilt of the otherwise horizontal resonances. As we explain below, the enhanced conductance lines are the signatures of the zero-energy Shiba state. The presence of the current enhancement is striking, since in usual STM setups the Shiba wavefunction is observed only up to a distance of 10~nm. On the contrary, we observe the Shiba state more than 50~nm away from the quantum dot.

Remarkably, applying a magnetic field smaller than the critical field, the conductance enhancement significantly increases (see Fig.~\ref{fig2}b, being measured in 150~mT). The largest conductance peak we observe is of size $\Delta G_{\text{T}} (B=150~\text{mT})\simeq 0.1~G_0$, approximately 20 times larger than the zero magnetic field value.

In a 250~mT magnetic field, above the critical field of the superconductor, $B_c = 230$~mT, the resonances vanish (see Fig.~\ref{fig2}c), indicating that the origin of the signal is related to superconductivity. To further illustrate the strong dependence of the signal on the magnetic field, we present in panel c line cuts at a fixed \VB$\simeq -1.5$~V for the same magnetic fields as in the other panels.

	\begin{figure*}[tb]
	\begin{center}
	\includegraphics[width=0.8\textwidth]{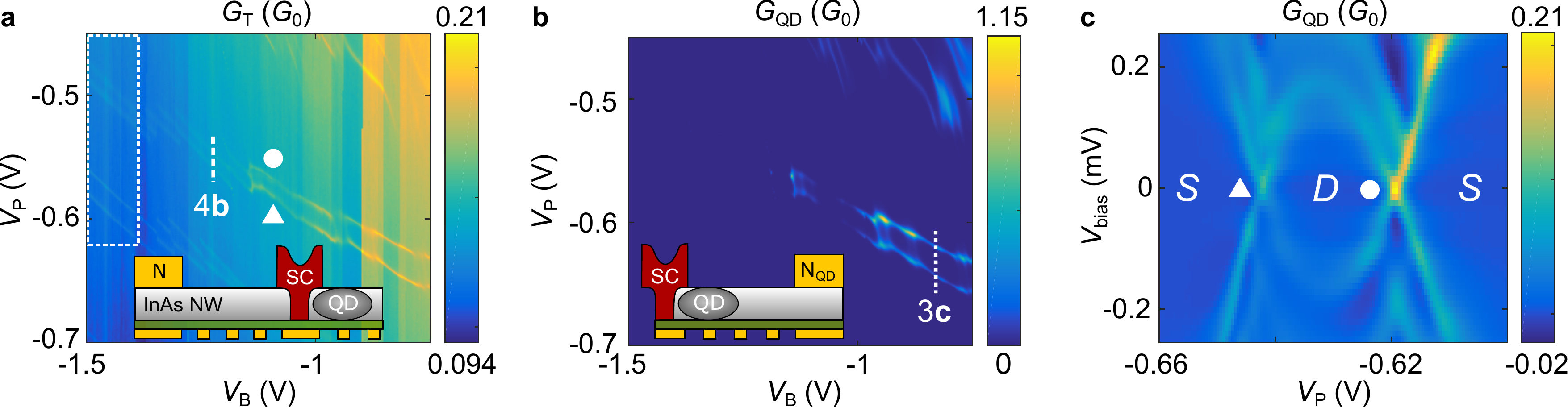}
	\caption{\textbf{Direct characterization of the Shiba state.} Differential conductance of the tunnel probe (panel \textbf{a}) and the quantum dot (panel \textbf{b}) in a larger gate voltage range without magnetic field. The white dashed rectangle marks the region already presented in Fig.~2a. For more positive $V_{\text{B}}$, the quantum dot also couples to the right normal lead, N$_{\text{QD}}$, while the conductance enhancement persists in $G_{\text{T}}$. The coincidence of the conductance lines on panels \textbf{a} and \textbf{b} indicates that the tunnel probe measures the level structure of the quantum dot non-locally. \textbf{c} Finite bias characterization of the quantum dot with open barrier at $V_{\text{B}}=-0.83$~V, along the white dotted line in panel \textbf{b}. The eye-shaped crossing is the signature of the Shiba state, $S$ and $D$ denotes the ground state character.}
	\label{fig3}
	\end{center}
	\end{figure*}

In the following, let us understand what condition of the quantum dot is linked to the tunnel current enhancement. As discussed in the Methods, our device is equipped with an extra normal electrode N$_{\rm QD}$ on the right of the quantum dot, which previously was isolated from the rest of the device by the large negative \VB voltage. Increasing $V_{\text{B}}$ to more positive values opens up the barrier to N$_{\text{QD}}$, which allows for a direct transport characterization of the quantum dot. In this way, we were able to measure in parallel both the differential conductance through the quantum dot itself, $G_{\rm QD}$, and the conductance through the tunnel probe, $G_{\text{T}}$. Figs.~3a and b show the conductance of the tunnel probe and the quantum dot, respectively, in a larger gate voltage window in the absence of an external magnetic field. The region marked by a white dotted rectangle is the particular voltage window in Fig.~\ref{fig2}a. Let us follow the resonances (marked by circle and triangle) in the plot of the tunnel current as $V_{\text{B}}$ increases.

For $V_{\rm B}\gtrsim -1.1$~V the tunnel barrier becomes sufficiently small, and transport through the quantum dot also sets in (see panel b of Fig.~\ref{fig3}). The similarities between the resonances in \GT and those in $G_{\rm QD}$ indicate that the two conductances are related; the tunnel conductance enhancement is linked to the level position of the quantum dot, i.e. the enhancement is observed when the dot is on resonance with the Fermi level of the superconductor.

To gain further insight to the level structure of the quantum dot, a finite bias measurement was performed along the white dashed line on Fig.~3b, at $V_{\text{B}}=-0.83$~V. The results are shown in Fig.~\ref{fig3}c. The eye-shaped crossing of the sub-gap conductance lines are the usual fingerprints of the Shiba state (see e.g. Ref.~\cite{EichlerPRL2007}). The results presented in Fig.~\ref{fig3} show that there is strong coupling between superconductor and quantum dot. In the previously shown measurements of Figs.~\ref{fig2}, \ref{fig3}a and \ref{fig3}b, the enhanced conductance lines correspond to the Shiba state when its energy is tuned to zero by \VP. These resonances (marked by white triangle and circle) correspond to the \textit{singlet-doublet} and \textit{doublet-singlet} transitions of the Shiba state. Since the position of the $G_{\text{T}}$ enhancement lines coincides with these transitions, we conclude that -- even in the case of large tunnel barriers when the quantum dot is coupled only to the superconductor (i.e. for $V_{\text{B}}<-1.1$~V) -- the conductance enhancement takes place when the energy of the Shiba state is tuned to zero. These results provide direct evidence that the tunneling electrode N indeed probes the Shiba state, and implies that the Shiba state extends in real space over the distance between the dot and the tunnel probe, separated by an impressive distance of $50-250$~nm. Here the width of the FIB cut gives the lower and the entire width of the superconducting lead the upper bound (see Fig.~\ref{fig1}b). 

\vspace{1cm}
\textbf{Further increased extension in magnetic field.}
A detailed analysis of the finite magnetic field behavior is presented on Fig.~4. Panel a shows the reduction of the superconducting gap, $\Delta$ with the magnetic field measured on the quantum dot. The gap smoothly decreases in magnetic field and continuously vanishes at $B_c=230$~mT, the critical field. The white dashed line is a fit, discussed below.

\begin{figure}[tb]
	\begin{center}
	\includegraphics[width=\columnwidth]{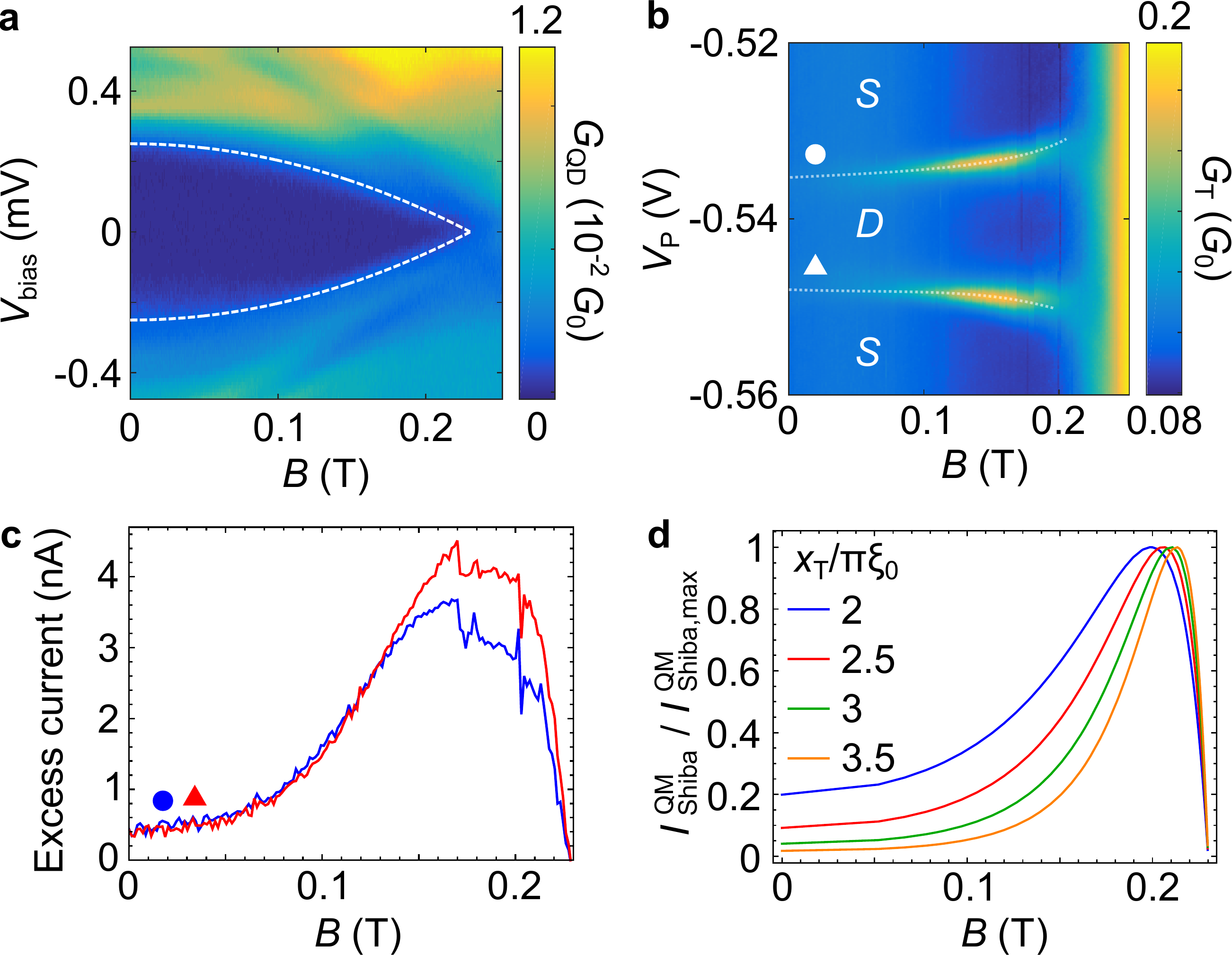}
	\caption{\textbf{Giant increase of the tunnel conductance in magnetic field.} \textbf{a} Reduction of the superconducting gap with increasing magnetic field measured at $V_{\text{P}}=-0.7$~V, $V_{\text{B}}=-0.6$~V. The dashed line shows the fit with Eq.~\eqref{eq:Delta-B}. \textbf{b} Evolution of the conductance enhancement with magnetic field measured along the dashed line on Fig.~3a. The gray dotted lines are guides to the eye. \textbf{c} The excess current (spectral weight) associated with the zero-energy Shiba state on panel \textbf{b}. The curves show the integrated excess tunnel conductance, $\Delta G_{\text{T}}(V_{\text{P}})$, after subtracting the $V_{\text{P}}$ independent background conductance. For the red/blue curve the integration along $V_{\text{P}}$ was carried out for resonance marked by triangle/circle on panel \textbf{b}. The integration window was converted to energy by the level arm of the dot. \textbf{d} NRG simulation of the excess current in the tunnel probe for different ratios of the distance between the dot and the tunnel probe, $x_{\text{T}}$ and the zero field coherence length, $\xi_0$ using $\varepsilon_d = -U/10$.}
	\label{fig4}
	\end{center}
	\end{figure}

The detailed evolution of the tunnel current enhancement with the magnetic field -- measured along the dashed line in Fig.~\ref{fig3}a at $V_{\text{B}}=-1.25$~V -- is shown on Fig.~\ref{fig4}b. While close to $B=0$, the peaks are barely visible, they are strongly enhanced with increasing magnetic field, particularly between 100 and 200~mT. For higher field values the peaks decrease and they disappear above $B_c$. 

\vspace{1cm}
\textbf{Discussion}

To explain the magnetic field dependence of the conductance enhancement and to probe the spatial extension of the Shiba state, we have set up a theoretical framework that allows us to compute the tunneling current though the normal lead N in a N--SC--QD geometry in a non-perturbative fashion. We assume that the quantum dot is coupled to the superconductor at $\bx=0$, while the normal lead is contacted to the superconductor further away at a coordinate ${\bf x}_{\text{T}}$. Moreover, we consider that the tunnel probe acts as an STM tip and measures the local density of states by injecting electrons at ${\bf x}_{\text{T}}$. Electrons entering the superconductor propagate to the quantum dot, scatter on it, and propagate back to be extracted at a later time but at the same position (The model is detailed in the Methods).

For a practical calculation, we need to determine the $T$-matrix that describes the scattering of the conduction electrons on the artificial atom. In our model, this is related to the Green's function of the creation operators on the quantum dot, as first discussed by Langreth~\cite{Langreth.1966}. Close to the parity changing transition and close to zero bias, the quasiparticles' contribution is irrelevant, and only the subgap states contribute. Using field theoretical methods, we computed the total current flowing from the normal lead through the Shiba state by performing a numerical renormalization group (NRG) calculation. 

Fig.~4c and d compare the experimental and NRG results as a function of magnetic field. In panel c the spectral weight for the experimental data is evaluated along the two dashed lines in panel b. The $V_{\text{P}}$ independent conductance background was subtracted from $G_{\text{T}}(V_{\text{P}})$, and the excess tunnel conductance $\Delta G_{\text{T}}(V_\text{P})$ was integrated  for the enhancement peaks (marked by circle and triangle) to get the total excess current/spectral weight induced by the Shiba state. The NRG-computed excess currents (see Eq.~\eqref{eq:current_NRG} of Methods) are displayed for different ratios of $x_{\text{T}}$ and $\xi_0$ in panel d, and show a close resemblance to the experimentally observed field dependence. For both panels and for low magnetic fields, the excess current strongly increases with the magnetic field, has a maximum, and linearly decreases at higher fields to vanish at the critical field, $B_c=230$~mT.

In order to understand this field dependence, it is instructive to display the results obtained for a classical spin on the dot (see Methods). This minimal model captures most of the experimentally observed features and it is in good agreement with the NRG results (see Methods). In the classical case, the current carried by the Shiba state at the transition reads    
\begin{equation}
I^{\text{Cl}}_{\text{Shiba}} = \frac{e}{h} \, {g_{\rm NS}}\,(|u_{\text{S}}({\bf x}_\text{T})|^2 + |v_{\text{S}}({\bf x}_\text{T})|^2) \,/ \,{\varrho_{\text{S}}} .
\end{equation} 
Here  $g_{\rm NS}$ stands for  the dimensionless conductance of the N--SC contact in the normal state (in units of $2e^2/h$), $u_{\text{S}}({\bf x}_\text{T})$ and $v_{\text{S}}({\bf x}_\text{T})$ denote the electron and hole parts of the bound Shiba state's wave function, and $\varrho_{\text{S}}$ is the density of states in the superconductor. The amplitude $|u_{\text{S}}({\bf x}_\text{T})|^2 + |v_{\text{S}}({\bf x}_\text{T})|^2$ can be easily computed for a spherical Fermi surface, yielding 
\begin{equation}
I^{\text{Cl}}_{\text{Shiba}} = e \;g_{\rm NS} \; {v_{\text{F}}\over 2\pi \xi} \;  e^{-{2 x_{\text{T}}}/{ (\pi\xi)}} \; \left| A({\bf x}_\text{T}) 
\right |^2,
\label{eq:current_main} 
\end{equation}
with $A({\bf x}_\text{T})$ a geometry, position, and  spatial dimension dependent dimensionless amplitude, and $\xi =\hbar v_{\text{F}}/\pi \Delta$ the superconducting coherence length with $v_{\text{F}}$ the Fermi velocity.
 
Apparently, the magnetic field dependence of the gap and thus that of the correlation length, $\xi_0\to \xi(B)$, is mostly responsible for the unusual magnetic field dependence observed in our experiment. In the framework of Ginzburg-Landau theory, the order parameter in an external magnetic field is given by $\Delta(B) = \Delta_0(1-B^2/B_c^2)^{1/2}$, where $\Delta_0\approx 0.25\, \text{meV} $ is the zero-field gap, $B$ the magnetic field, and $B_c$ stands for the critical magnetic field at which superconductivity vanishes. Experimentally, however, we find a slightly different functional form for the suppression, qualitatively similar to that observed in thin films~\cite{TinkhamPRL1961,AnthorePRL2003} (dashed line on Fig.~4a), as
\bnen 
\label{eq:Delta-B} \Delta(B) \approx \Delta_0 \left( 1 - \frac{B^2}{B^2_c} \right), 
\eden
with $\Delta_0 = 250~\mu$eV  and $B_c = 230$~mT. For $x_{\text{T}}> \xi_0$, Eq.~\eqref{eq:Delta-B} together with Eq.~\eqref{eq:current_main} implies an exponential increase in the current with increasing magnetic field and then a suppression close to $B_c$ due to the prefactor $\xi^{-1}$. This is consistent with the upturn of measured current enhancement at low fields, below $120$~mT (see Fig.~\ref{fig4}c), and its suppression close to the critical field. According to Eq.~\eqref{eq:current_main}, the current should be maximal for $x_{\text{T}}\approx \pi \xi(B) /2$. Using $B \approx 180$~mT -- where the total excess current is maximal (see Fig.~\ref{fig4}c) -- this condition yields to $x_{\rm T}\approx 3 \xi_0$. However note that the theory slightly overestimates the magnetic field value where the excess current is maximal -- and so underestimates the ratio of $x_{\text{T}}/\xi_0$. This deviance may originate from the fact that our model idealizes the setup, i.e. neglects the presence of quasiparticles and a possibly finite subgap density of states. Keeping this deviance in mind, the obtained ratio is still roughly consistent with our geometrical parameters, the separation of the quantum dot and the normal electrodes $x_{\text{T}} \approx 50-250 \,\text{nm}$, and a reduced coherence length, $\xi_0 \approx 20-40 \,\text{nm}$, compatible with a diffusive superconductor.

Let us turn to differences compared to typical STM measurements of Shiba states: In STM characterization of the Shiba wavefunction, $I_{\text{Shiba}}$ is significantly weaker. Moreover, it oscillates upon varying the tip-adatom distance, $\bx_{\text{T}}$, and becomes unobservable at distances larger than 10~nm for 3D superconductors. These characteristic differences originate in the quite different geometries used in our setup and in STM experiments.
 
In STM measurements, the amplitude $|A(\bx_{\text{T}})|^2$ is responsible for the spatial oscillations. It has a modulation of $~|\sin(k_{\text{F}} x_{\text{T}})|^2$, which originates in the point-like nature of the tunnel probe \cite{BalatskyRMP2006,MenardNatPhys2015}, and the fact that one usually tunnels either to the electron- or hole-like states with amplitudes $|u_{\text{S}}({\bf x}_\text{T})|^2$ and $ |v_{\text{S}}({\bf x}_\text{T})|^2$, respectively. In contrary, in our setup, two effects seem to eliminate these oscillations. First, the N--SC interface has a large tunnel surface ($\approx d^2$ where $d$ is the nanowire diameter, $d \approx 80$~nm), and averaging for different distances and tunneling paths is expected to eliminate oscillations both in distance (not accessible in our setup) as well as in the $B$ field dependence \cite{Bouchiat2005}. Second, at the transition point the electron and hole-like contributions add up and, surprisingly, the the combination  $|u_{\text{S}}({\bf x}_\text{T})|^2 + |v_{\text{S}}({\bf x}_\text{T})|^2$ does not contain {\it any} oscillating term in any dimension. This indicates that interference effects probably play little role right at the transition, where our measurements are carried out.   

In an STM geometry $I_{\text{Shiba}}$ is a small signal ($\sim$pA) \cite{MenardNatPhys2015}, while in our measurement, it can reach values as large as $I_\text{Shiba,max}\approx 0.7 \; \text{nA}$ (at the excitation of $V_{\text{AC}}=10~\mu$V), even though the normal electrode is at a separation  $x_{\text{T}} \approx 50-250$~nm away from the quantum dot. According to Eq.~\eqref{eq:current_main}, $I_{\text{Shiba}}$ is directly proportional to $g_\text{NS}\; |A|^2$. The small $I_{\text{Shiba}}$ for typical STM experiments is a result of two factors: the weak tunnel coupling $g_{\rm NS}\ll 1$ and also the fast, $1/ (k_{\text{F}} x_{\text{T}})^2$  decay of $|A(\bx_{\text{T}})|^2$ in a 3D superconductor. In our case, $g_\text{NS}$ is relatively large, in the range of $0.1-1$. Additionally, the N--SC tunneling area is very large, $\sim d^2\sim x_{\text{T}}^2$, compensating the decay of $|A(\bx_{\text{T}})|^2$, in close resemblance to lower dimensional superconductors. A combination of these effects can explain how the Shiba state can be observed even from a remote position, significantly larger than the one in STM measurements \cite{JiPRL2008,ChoiNatComm2017,MenardNatPhys2015,RubyPRL2016}. In our quantum dot based setup, the observation distance is determined essentially only by the coherence length of the superconductor, and the coupling between Shiba states hosted around quantum dots can be achieved at significantly larger distances than in case of adatoms.

Finally let us contrast our findings with the results of Cooper pair splitter measurements, where two quantum dots are attached to both sides of the superconductor, and contrary to our setup, current flows from the superconductor through both quantum dots towards the normal leads. Current correlations between the two arms are induced by splitting up Cooper pairs \cite{HofstetterNature2009,HermannPRL2010,HofstetterPRL2011,DasNatComm2012,SchindelePRL2012}. There, electronic correlations extend over distances of about $50-200$~nm, comparable with the separation $x_{\text{T}}\approx 100-250$~nm in our experiment. However, while the Shiba current, $I_{\text{Shiba}}$ increases in a magnetic field, the Cooper pair splitting signal gets strongly suppressed \cite{HofstetterPhD2011}.

In conclusion we have studied the Shiba state in a SC--QD hybrid device by measuring the differential conductance in a tunnel probe attached to the superconductor at distance of $50-250$~nm away from the quantum dot. A large current enhancement has been observed when the Shiba resonance is tuned to zero energy. In an external magnetic field, the signal is further enhanced, implying an exponential growth for the extension of the Shiba state. The observed behavior is consistent with our  microscopic theoretical model and field theoretical calculations. In our device, we can access the Shiba state from a remarkably large distance compared to previous experiments on magnetic impurities on superconducting substrates. These results establish an important milestone towards the realization of Shiba-chains implemented by a series of quantum dots  attached to superconductors.

\vskip1cm
{\bf \noindent \large Acknowledgments}

We acknowledge Morten~H.~Madsen for MBE growth, Jens~Paaske, Andr\'as~P\'alyi, Dimitri~Roditchev and Pascal~Simon for useful discussions. We also acknowledge SNI NanoImaging Lab for FIB cutting.

This work was supported by the National Research Development and Innovation Office of Hungary within the Quantum Technology National Excellence Program 2017-1.2.1-NKP-2017-00001, by the New National Excellence Program of the Ministry of Human Capacities, by QuantERA SuperTop project 127900, by AndQC FetOpen project, by Nanocohybri COST Action CA16218, and by the Danish National Research Foundation. CPM was supported by the Romanian National Authority for Scientific Research and Innovation, UEFISCDI, under project no.~PN-III-P4-ID-PCE-2016-0032. CS acknowledges support from the Swiss National Science Foundation through grant Nr.~20020$\_$L7263a and through the National Center of Competence in Research on Quantum Science and Technology (QSIT).

\vskip1cm
{\bf \noindent \large Author contributions:}

Experiments were designed by C.~S. and S.~C., wires were developed by J.~N., devices were developed and fabricated by G.~F. and J.~G., measurements were carried out and analyzed by Z.~S., G.~F., A.~B., P.~M., T.~E. and S.~C. Theoretical analysis was given by C.~P.~M. and G.~Z.. And Z.~S., C.~P.~M., G.~Z. and S.~C. prepared the manuscript. 

\vskip1cm
{\bf \noindent \large Additional information:}
The authors declare no competing interests. Correspondence should be addressed to S.~C..

\vskip1cm
{\bf \noindent \large Methods}

\vspace*{5mm}
{\bf\noindent Sample fabrication and measurement details.} 
\vspace*{2mm}

An SEM micrograph of the measured device is shown in Fig.~5. First, 9 bottom gate electrodes were defined by electron beam lithography and evaporation of 4~nm Ti and 18~nm Pt. Two $1.3~\mu$m wide gates were used below the normal contacts and one, 250~nm wide below the superconductor. The remaining 3+3 bottom gates are 30~nm wide with 100~nm period. The gates were covered by 25~nm SiN$_\text{x}$, using plasma enhanced chemical vapor deposition, to serve as an insulating layer \cite{FulopPRB2014}. The SiN$_\text{x}$ was removed at the end of bottom gate electrodes by reactive ion etching with CHF$_3$/O$_2$ \cite{WongJVacSciTechB1992}, to contact the gate electrodes. The InAs nanowire was deposited by micro-manipulator onto the SiN$_\text{x}$ layer, approximately perpendicular to the bottom gates. The NWs were grown by gold catalyst assisted MBE growth \cite{AagesenNatNanotech2007,MadsenJCrystGrow2013}, using a two-step growth method to suppress the stacking faults \cite{ShtrikmanNanoLett2009}. The normal (N$_{\text{L}}$ and N$_{\text{R}}$), Ti/Au (4.5/100~nm) and superconducting (SC), Pd/Pb/In (4.5/110/20~nm) contact were defined in further e-beam lithography and evaporation steps \cite{GramichAPL2016}. The later has a width of 250~nm. Prior to the evaporation, the nanowire was passivated in ammonium sulfide solution to remove the native oxide from the surface \cite{SuyatinNanoTech2007}.

	\begin{figure}[tb]
	\begin{center}
	\includegraphics[width=0.75\columnwidth]{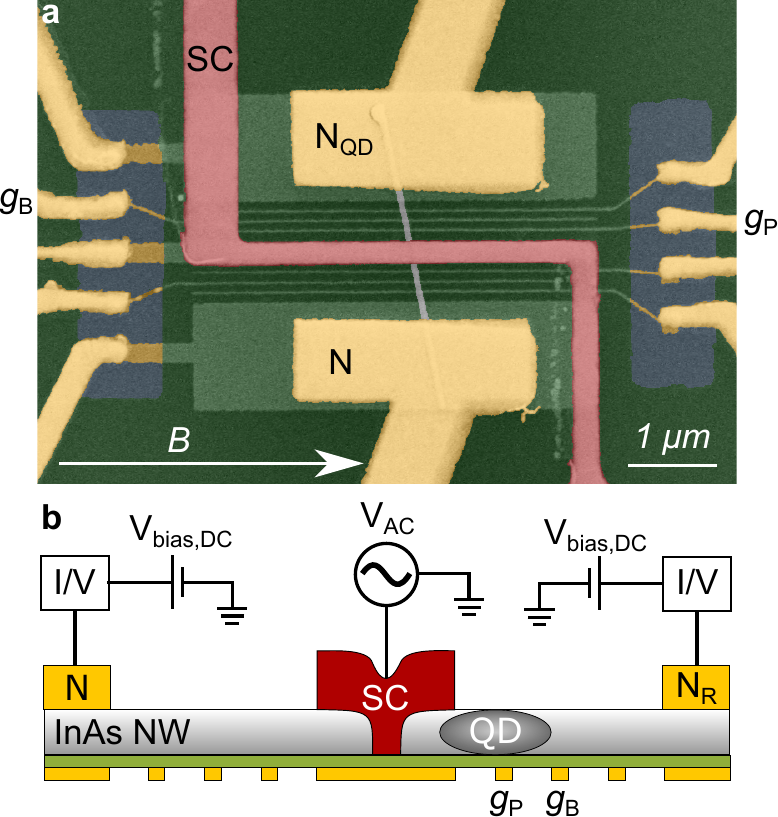}
	\caption{Device geometry. \textbf{a} False color SEM image of the measured device. \textbf{b} Schematics of the cross-section of the device together with the circuit diagram.}
	\label{fig5}
	\end{center}
	\end{figure}

Prior to the deposition of the superconducting contact the nanowire was cut by FIB to prevent direct tunneling between the NW segment, which can lead to spurious effects \cite{FulopPRB2014,FulopPRL2015}. The width of the FIB cut was about 50~nm, giving a lower bound for the distance of the quantum dot and the tunnel probe.

The measurements were done in a Leiden Cryogenics CF-400 top loading cryo-free dilution refrigerator equipped with a 9+3~T 2D vector-magnet. The measurements were done at a bath temperature of 35~mK. Prior to the cool down, the sample was pumped overnight to remove the adsorbed water contamination from the surface of the nanowire. The currents were measured by standard lock-in technique at 237~Hz. The AC signal of $V_{\text{AC}}=10~\mu$V was applied to the superconducting electrode. The currents in the left and right arm were measured simultaneously via the two normal leads by home-built I/V converters. The DC bias was applied symmetrically to the normal leads. An in-plane magnetic field was applied parallel to the superconducting electrode, perpendicular to the nanowire. The circuit diagram is shown in Fig.~5b. In most of the measurements presented in the main text, N$_{\text{QD}}$ was electrically isolated from the rest of the device by applying a large negative voltage on gate $g_{\text{B}}$.

\vspace*{5mm}
{\bf\noindent Shiba bound state and the parity crossing transition.}	

We model the superconductor using the s-wave BCS theory~\cite{Schrieffer}
\begin{equation}
H_{\rm SC} = \sum_{\bk \s} \varepsilon_\bk c^\dagger_{\bk\s}c_{\bk\s} +\Delta \left(c^{\dagger}_{\bk\uparrow} c^\dagger_{-\bk \downarrow} +h.c. \right),
\label{eq:BCS}
\end{equation}
where $c^\dagger_{\bk\s}$ is the creation operator of a spin $\s$ and momentum $\bk$ and $\Delta$ is the superconducting order parameter, which can be considered real. 

\begin{figure}[tb]
	\begin{center}
	\includegraphics[width=0.7 \columnwidth]{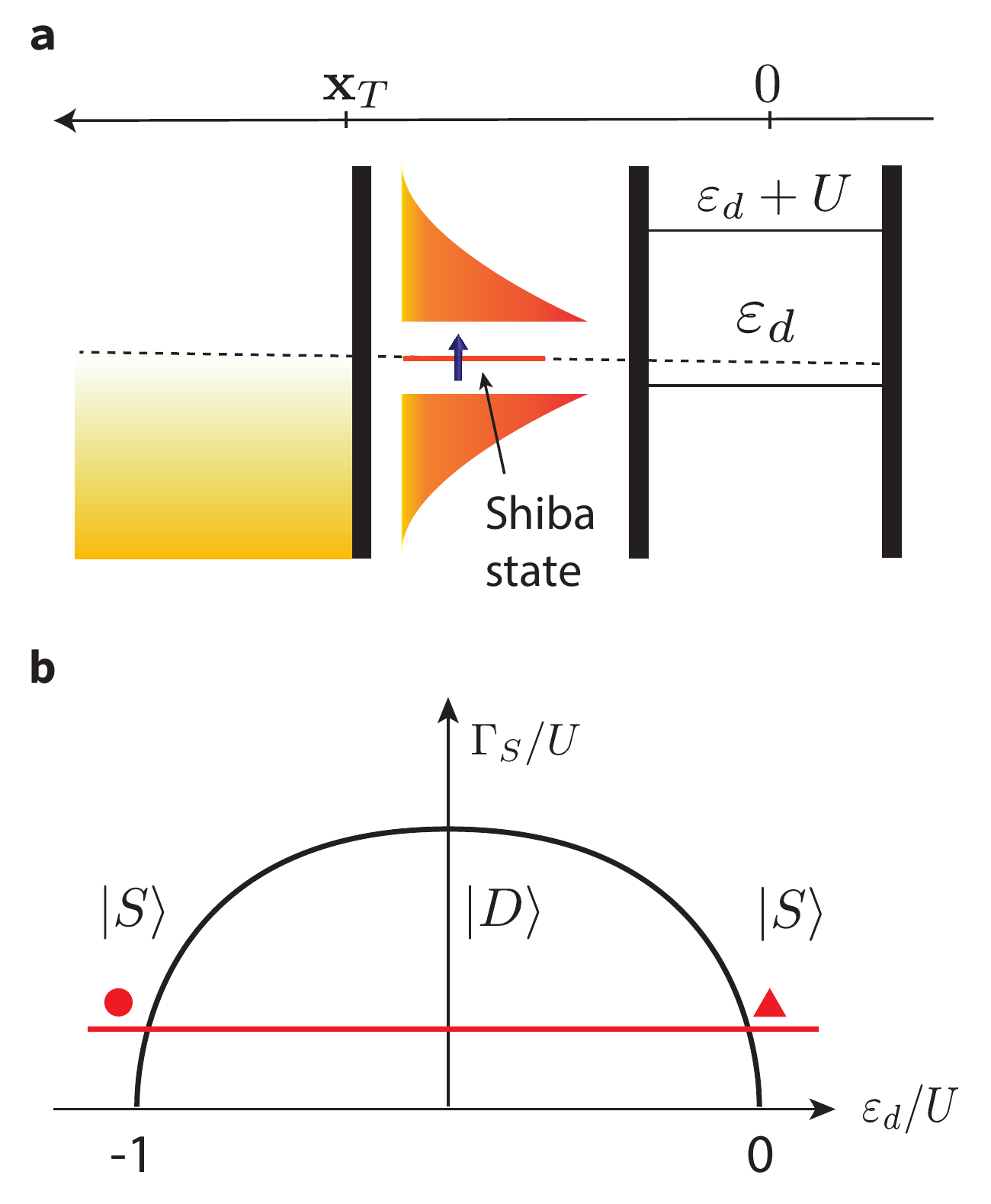}
	\caption{Phase diagram. \textbf{a} Schematics with all the energies involved. \textbf{b} The typical phase diagram. The dome is the boundary for the stability of the magnetic doublet state (inside) versus the singlet ground state (outside).}
	\label{fig:phase_diagram}
	\end{center}
	\end{figure}

The QD can be described by means of the Anderson model, 
\begin{equation}
H_{\rm QD} = \sum_{\s}\varepsilon_d d^\dagger_{\s}d_{\s} +U n_{\uparrow}n_{\downarrow}, 
\label{eq:H_QD}
\end{equation}
with $\varepsilon_d$ is the single particle energy and $U$ is the on-site Coulomb energy. The operator $d^\dagger_\s$ here creates an electron with spin $\s$ on the dot, and $n_\s = d^\dagger_\s d_\s$. In our geometry the dot is located at the position $\bx=0$ and is tunnel-coupled to the superconductor (See Fig.~\ref{fig:phase_diagram}a), as described by the Hamiltonian
\begin{equation}
H_{\rm SC-QD} = t_{\text{S}}\sum_{\s,\bk}\left ( \alpha_\bk \, c^\dagger_{\bk \s} d_{\s}+h.c. \right) .
\label{eq:H_tun}
\end{equation} 
The factor $\alpha_\bk$, normalized to one at the Fermi surface, $\langle |\alpha_\bk|^2\rangle_{\text{F.S.}} =1$, accounts for the directional dependence of the tunneling, and prefers tunneling directions perpendicular to the SC--QD interface in our geometry.  

In the local moment regime, we can neglect charge fluctuations of the dot, and describe it in terms of a simple Kondo model, 
\begin{equation}
H_{\rm imp} = {1\over 2}\sum_{\bk, \bk'}J_{\bk\bk'} \bS\, c^\dagger_{\bk\s} \bs_{\s\s'}c_{\bk'\s'},
\label{eq:exchange}
\end{equation}
with $J_{\bk\bk'} = \alpha_{\bk}\alpha_{\bk'} J_0$, and $J_0 \propto t_{\text{S}}^2/U$ the exchange coupling to the spin of the quantum dot, $\bS$. 

This latter Hamiltonian can be solved exactly in the classical limit~\cite{ShibaProgTheorPhys1968}, where one finds a subgap resonance at an energy:
\begin{equation}
E_0 = \Delta \frac{1-\tilde j^2}{1+{\tilde j}^2},
\label{eq:energy}
\end{equation}
where $\tilde j = \pi j/2$, and $j = J_0\varrho_{\text{S}}$ stands for the standard dimensionless Kondo coupling. With increasing coupling strength the bound state energy $E_0$ eventually crosses zero, and the impurity binds to itself a quasiparticle of opposite spin direction. 
 
Although this classical calculation captures the parity changing transition, determines its location incorrectly. In reality, the phase transition originates from the competition between the superconducting correlations and the Kondo screening of the spin, and the transition takes place when $\Delta \simeq T^*$, with $T^*$ a characteristic Fermi liquid temperature scale. Deep in the local moment regime, $T^*$ can be identified as the Kondo temperature, $T^*\simeq T_{\rm K}$, but it becomes of the order of the tunneling rate $\Gamma_{\rm S}$ between the dot and the superconductor close to the mixed valence regime. For $T^*<\Delta$, the  spin of  the dot remains unscreened down to zero temperature, resulting in a doublet ground state \db=\{$| \Uparrow\rangle, |\Downarrow\rangle$\}  and a first excited singlet state, \sg, the so-called Shiba state inside the gap. By increasing the tunneling rate \GS, the spin in the QD binds a quasiparticle from the superconductor, and \sg  becomes the ground state. The corresponding qualitative phase diagram is displayed in Fig.~\ref{fig:phase_diagram}b.

In the experiments, changing the voltage $ V_{\rm P}$ corresponds to moving along the red horizontal line in the phase diagram. At the intersections with the "dome" (denoted by the circle and the triangle), the Shiba state is at zero energy, and is in resonance with the Fermi energy, $E_{\text{F}}$ of the tunneling electrode N, leading to the observed  sudden increase in the zero voltage differential conductance. 

\vspace*{5mm}
{\bf\noindent Tunnel conductance and current.}

In our setup, the normal lead acts as an STM tip measuring the differential conductance across the device at a point $\bx_{\text{T}}$ away from the dot. In the small SC--N tunneling limit Andreev processes can be neglected, and tunneling from the normal electrode to the superconductor yields at $T=0$ temperature a current
\begin{equation}
I = {e\over h}\int_{0}^{eV} |t_{\text{N}}|^2 d\omega \, \varrho_{\text{N}} \,\varrho(\bx_{\text{T}},\omega),
\label{eq:current}
\end{equation}
at a finite voltage bias $V_{\rm bias} = V$. Here $\varrho_{\text{N}}$ denotes the density of states in the normal lead, $t_{\text{N}}$ is the tunneling amplitude between the superconductor and the normal contact, and $\varrho(\bx, \omega) = - \text{Im}\, G_\s(\bx, \omega)/\pi$ is the energy dependent density of states in the superconductor at position $\bx_{\text{T}}$. This latter can be expressed in terms of the Fourier transform of the retarded Green's function, $G_\sigma(\bx_{\text{T}},\omega) = -i \langle \{ \psi_\s(\bx_{\text{T}},t) ,\psi^\dagger_\s(\bx_{\text{T}},0) \} \rangle\,\Theta(t)$.
 
The Green's function $G_\sigma(\bx_{\text{T}}, \omega)$ can be obtained by means of standard many-body theory, invoking Nambu spinors, $\{\phi_{\sigma\tau}\} \equiv (\psi_\uparrow,\psi_\downarrow, \psi^\dagger_\downarrow,-\psi^\dagger_\uparrow)$, and corresponding propagators, $G\to \hat G$. In this language, the  quantum dot induced  part of the propagator $\hat G$ can be expressed as 
\begin{equation}
\delta \hat G(\bx_{\text{T}}, \omega) = -  \hat g(\bx_{\text{T}}, 0, \omega) \hat{\cal T}(\omega) \,\hat  g(0, \bx_{\text{T}}, \omega).
\end{equation}
Here $\hat g(\bx_{\text{T}}, 0, \omega)$ describes the propagation of an electron (or hole) in the superconductor from the dot to the probe at an energy $\omega$, 
\begin{equation}
\hat g(\bx_{\text{T}}, 0, \omega) = \int \frac{{\rm d}^3\bk}{(2\pi)^3}\frac { \alpha_\bk \, e^{i \bk \cdot \bx_{\text{T}}} }{\omega + i \delta - (\xi(\bk) \tau_z + \Delta \tau_x)},
\label{propagator}
\end{equation}
with the Pauli matrices $\tau_x$ and $\tau_z$ acting in the Nambu space, while the $T$-matrix $\hat {\cal T}(\omega)$ describes scattering off the quantum dot.  

We focus on transport through midgap states, i.e. $|\omega| < \Delta$. There the propagators $ \hat g(\bx_{\text{T}}, 0, \omega)$ are real, and the tunneling density of states $\varrho(\bx, \omega)$ is related to ${\rm Im} \hat{\cal T}(\omega)$.

In the classical limit, we can determine $\hat {\cal T}$ analytically,
$$
\varrho_{\text{S}} \hat {\cal T}(\omega)  = \frac{j \bS \cdot \bs \sqrt{\Delta^2-\omega^2}} {\sqrt{\Delta^2-\omega^2} + j \pi \bS \cdot \bs (\omega + \Delta \tau_x)}
$$
and extract the strength of its poles within the gap. At the transition points, $j = 2/\pi$, and expressing furthermore propagator $\hat g(\bx_{\text{T}}, 0, \omega=0)$ we arrive at the equation~\eqref{eq:current_main} displayed in the main text. 
 
In the framework of the Anderson model, we first introduce the spinors $\{D_{\sigma\tau}\} \equiv (d_\uparrow,d_\downarrow, -d^\dagger_\downarrow, d^\dagger_\uparrow)$. Similar to \cite{Costi}, it is easy to show that the $T$-matrix is directly related to the $d$-levels' Nambu propagator, 
$$
\varrho_{\text{S}} \hat {\cal T}(\omega)  = \varrho_{\text{S}}  \; t_{\text{S}}^2 \; \hat {\cal G}_{DD}(\omega)\;.
$$
The subgap structure of $\hat {\cal G}_{DD}$ is completely determined by the transition matrix elements $\alpha \equiv \langle \Uparrow| d^\dagger_\uparrow| S\rangle$ and $\beta \equiv \langle \Uparrow| d_\downarrow| S\rangle$, and the energy difference $E_S$ of the singlet ground state $|S\rangle$ and the many-body Shiba excitations, $|\Uparrow\rangle$ and $|\Downarrow\rangle$, related by time reversal symmetry. At the transition point, the electron and hole contributions  add up, and the expression of the tunnel current simplifies to
\begin{equation}
I^{\text{QM}}_{\text{Shiba}}  = \frac{e}{h}  \;g_{\rm NS}  \;  \pi\, \Gamma_{\text{S}} (|\alpha|^2 +|\beta|^2)\;
 e^{-{2 x_{\text{T}}}/{ (\pi\xi)}} \; \left| A({\bf x}_\text{T}) 
\right |^2 ,
\label{eq:current_NRG}
\end{equation}
with $\Gamma_{\text{S}} = 2\pi t_{\text{S}}^2 \varrho_{\text{S}}$, and $\Gamma_{\text{S}}/\hbar$ the tunneling rate from the quantum dot to the superconductor. The results shown on Fig.~\ref{fig4}d of the main text are generated by Eq.~\eqref{eq:current_NRG}. Notice that the many-body expression differs from the classical one solely by the factor $\Gamma_{\text{S}} (|\alpha|^2 +|\beta|^2)$, which replaces the gap $\Delta = \hbar v_{\text{F}}/(\xi\pi)$ in the prefactor of the classical expression. 

\begin{figure}[tb]
\begin{center}
\includegraphics[width=0.8\columnwidth]{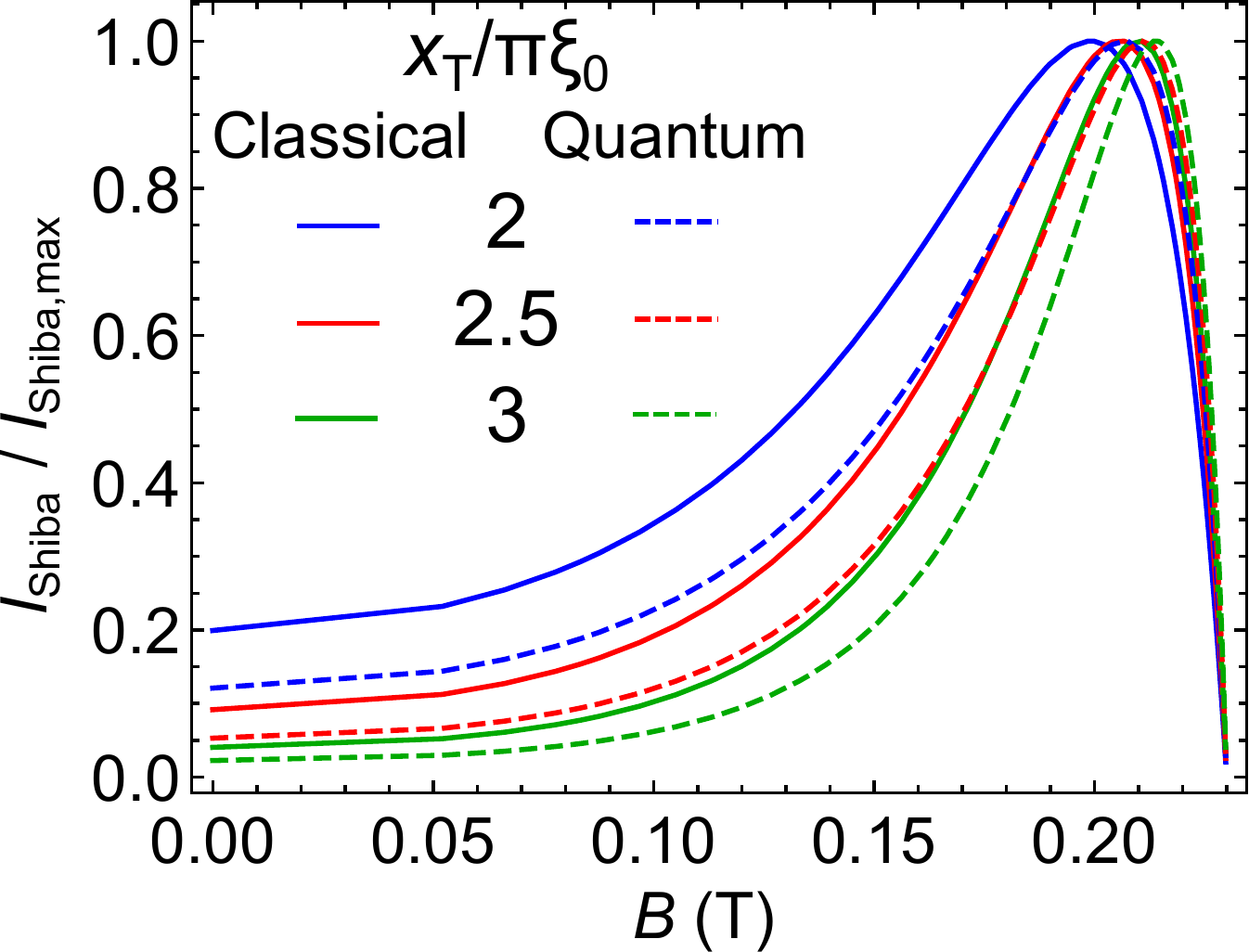}
\caption{Comparison between the classical expression, Eq.\eqref{eq:current_main} and result involving the NRG calculation, Eq.~\eqref{eq:current_NRG} for the current in the tunneling contact.}
\label{fig:comparison}
\end{center}
\end{figure}

The matrix elements $\alpha$ and $\beta$ can be directly extracted from NRG computations. A comparison between the two approaches  is presented in Fig.~\ref{fig:comparison}. They both give similar results. Small quantitative differences are presumably due to the fact that the classical picture is not able to capture neither the mixed valence regime nor the proper Kondo scale, in contrast to the more elaborate NRG approach. 

\vspace*{5mm}
{\bf\noindent Data availability} 
\vspace*{5mm}

The data that support the plots within this paper and other findings of this study are available at \url{https://doi.org/10.5281/zenodo.3604194}.

\bibliography{scheriff}

\end{document}